\documentclass[11pt]{article}
\usepackage{moriond,epsfig}

\def\Journal#1#2#3#4{{#1} {\bf #2}, #3 (#4)}

\def\PRD{{\em Phys. Rev.} D}

\newcommand{\apj}[0]{ApJ}
\newcommand{\aap}[0]{A\&A}
\newcommand{\mnras}[0]{MNRAS}

\def\be{\begin{equation}}
\def\ee{\end{equation}}
\def\bea{\begin{eqnarray}}
\def\eea{\end{eqnarray}}

\begin{document}
\vspace*{4cm}
\title{JETS DRIVEN BY ACCRETION ONTO SPINNING BLACK HOLES}

\author{ IOANA DU\c{T}AN~$^{1,2}$ and PETER L. BIERMANN~$^{1,3}$ }
\address{$^1$ Max-Planck-Institut f\"ur Radioastronomie, Auf dem H\"ugel 69,
  D-53121 Bonn, Germany\\
$^2$ University of Bucharest, Faculty of Physics, Bucharest-Magurele, Romania\\ 
$^3$ University of Bonn, Dept of Astronomy \& Physics, Germany
}

\maketitle\abstracts{
We present a theoretical model for driving jets by accretion onto Kerr black holes and try to give an answer to the following question: \textit{How much energy could be extracted from a rotating black hole and its accretion disk in order to power relativistic jets in Active Galactic Nuclei?} }

\section{Introduction}
Recent developments of astronomical observations have revealed that the universe is full of powerful phenomena, such as jets. The relativistic jets emerging from Active Galactic Nuclei (AGNs) -- Quasars, Radio Galaxies, Blazars, Seyfert galaxies -- are the biggest in size propagating through the intergalactic medium and the most energetic of the whole range of astrophysical jets. Some of them may be associated with highly rotating, supermassive black holes (of mass up to $M\approx 10^9\,M_{\odot}$) surrounded by an accretion disk and harboured in the center of a radio-loud Radio Galaxy. While the accretion disk shows thermal emission (for low accretion rates, very weak), in contrast, the emission from jets is typically non-thermal in a large frequency range that can go from radio to $\gamma$-ray bands; some of them are connected to an accretion disk with a sub-Eddington accretion rate (see, e.g., Yuan {\it et al.}~\cite{yuan}; Falcke {\it et al.}~\cite{fkm}). This kind of jet cannot be driven by just a purely standard relativistic accretion disk (Shakura \& Sunyaev\cite{ss}; Novikov \& Thorne \cite{nt}, hereafter NT73); the gravitational energy released from the accretion disk for driving jets could be increased by the rotational energy of the black hole (BH) extracted via magnetic fields. The energy and angular momentum are taken from the BH through poloidal magnetic field lines that connect the accretion disk to the BH; i.e., through a magnetic coupling process (see, e.g., Li~\cite{li1}), which is a variant of the Blandford-Znajek mechanism~\cite{bz}.

\section{Total energy-flow along the jets} 
The Kerr metric describes the curved spacetime around a spinning BH, which has two characteristic parameters: mass $M$ and angular momentum $J$. For a steady, axisymmetric, and thin accretion disk (the thickness of the disk is much smaller than its radius) around a Kerr black hole, the general relativistic equations of angular momentum conservation and energy conservation have been investigated by Page \& Thorne~\cite{pt}. The particles from the accretion disk flow along direct, nearly circular, orbits in and near the equatorial plane of the BH. For any Kerr BH, there is an invisible boundary known as the static limit; at and inside that limit particles from the accretion disk are dragged in the same direction as the BH rotation. For the equatorial plane of a Kerr BH, the radius of the static limit is $r_{sl}=2r_g$ (with $r_g=GM/c^2$ being the gravitational radius). The angular momentum is transported by torques in the disk due to viscous stresses resulting in a radial flow of the particles towards the center. In the case of a thin accretion disk, when the particles reach the innermost stable orbit $r_{ms}$ they drop out of the disk and plunge into the black hole. That is to say, at $r_{ms}$ the torque vanishes. At present the torque from the magnetic fields is not yet included. In this picture the accretion disk is efficiently radiative (more details can be found in NT73). The disk properties depend on the spin parameter $a_*=a/r_g$ ($0\leq a_*\leq 1$); where $a=J/Mc$ (Macdonald \& Thorne~\cite{mt}; hereafter MT82) is the angular momentum of the BH per unit mass per speed of light about the spinning axis. 

Considering that the innermost region of the accretion disk extends from the static limit to the innermost stable orbit, in which case the dominant torque in the disk is due to the magnetic stresses produced by poloidal magnetic fields (that connect the accretion disk to the BH), and assuming that the jets are driven from this innermost region of the disk, the energy conservation law could be written as follows:
\begin{equation}
\frac{d}{dr}\left[ \left( 1-q_m\right) {\dot{M}}_{D}c^2E^{\dagger}\right] =4\pi r\left( JE^{\dagger}-H\Omega_{D}\right)\,.
\label{eq1}
\end{equation}
The first term describes the energy transported by the rest mass of the accreting gas, the second term describes the energy transported along the jets, and the last term describes the energy transported in the disk due to the magnetic stresses. Here the accretion disk is perfectly conducting and the magnetic field lines are frozen in the gas. 

For the above equation Eq.~\ref{eq1}, $q_m$ is the ratio of the mass rate flow along the jets ${\dot{M}}_{jets}$ to the accretion rate ${\dot{M}}_{D}$ (${\dot{M}}_{D}=\dot{m}{\dot{M}}_{Edd}$, with ${\dot{M}}_{Edd}$ being the Eddington accretion rate corresponding to an efficiency in converting the rest mass energy of the infalling material into radiation of $10\,\%$), $E^{\dagger}$ is the specific energy of the particles in the disk, $J$ is the energy-flow along the jets, $H$ is the flux of the angular momentum transported by the poloidal magnetic field (see Li~\cite{li2}), and $\Omega_{D}$ is the angular velocity of the accretion disk.

\begin{figure}
\begin{center}
\includegraphics[width=0.5\textwidth]{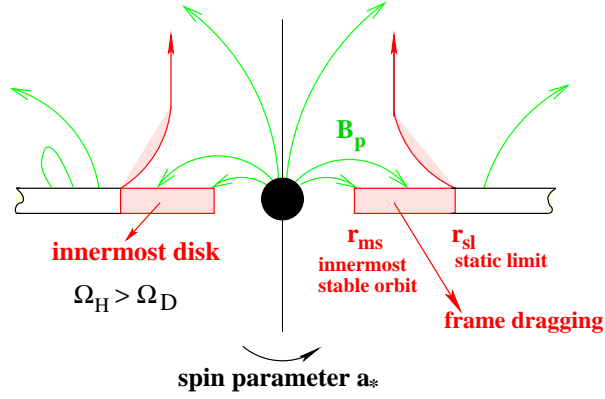}
\end{center}
\caption{Schematic representation of the innermost region of the accretion disk threaded by a poloidal magnetic field which connects BH, accretion disk, and jet.} 
\label{pol}
\end{figure}

The magnetic fields, as well as the accretion process (since there is always accretion), seem to play an important role in transporting energy along the jets. The minimum accretion is from a red giant star wind -- and there is always a red giant around. Of course, if the ram pressure of the wind is insufficient to reach the neighbourhood of the BH, then there is only Bondi-Hoyle accretion from the confining hot medium. The poloidal magnetic field lines could be closed at very large distances, but some of them may be closed in the innermost region of the accretion disk (Figure~\ref{pol}), such that the energy and angular momentum are transported from the BH to the disk via flux $H$.

Integrating Eq.~\ref{eq1} over the disk surface between the innermost stable orbit and the static limit, we obtain the total energy-flow (or power) along the jets 
\begin{equation}
P_{jets}=\left( 1-q_m\right) {\dot{M}}_{D}c^2\left[ E^{\dagger}\left( r_{sl}\right) -E^{\dagger}\left( r_{ms}\right) \right]+4\pi \int_{r_{ms}}^{r_{sl}}rH\Omega_{D}dr\,.
\end{equation}
Here the energy is extracted from the accretion disk (first term) and from the rotational energy of the black hole  via the accretion disk (second term). 

In order to make further progress, it is necessary to specify the flux of the angular momentum transported by the magnetic connection between the disk and the black hole
\begin{equation}
H=\frac{1}{8\pi^{3}r}\left( \frac{d\Psi_{D}}{c\,dr}\right) ^{2}\frac{\Omega_{H}-\Omega_{D}}{\left(-dR_{H}/dr\right) }\,,
\label{eq2}
\end{equation}
where $\Psi_{D}$ is the flux of the poloidal magnetic field which threads the accretion disk, $\Omega_{H}$ is the angular velocity of the Kerr BH, and $R_{H}$ is the effective resistance of the horizon, which is practically the resistance of free space $4\pi/c =377 \: \textrm{ohm}$ (see, e.g., MT82) at the end of an open waveguide since the horizon of BH behaves, in some aspects, like an electrically conducting surface. Such behavior comes from the fact that whenever a BH is in stationary equilibrium, the electric potential over its surface is constant (Carter~\cite{carter}). There is one more point that should be made here: the BH resistance as a function of the accretion disk radius $(-dR_{H}/dr)$ could be found by considering the continuum of the magnetic field within a narrow strip between two magnetic surfaces that connect the accretion disk to the BH and by using the relation between the strength of the magnetic field that threads the black hole $B_{H}$ and the poloidal magnetic field at the inner edge of the disk $B^{p}_{D}(r_{ms})$, $B_{H}=\zeta B^{p}_{D}(r_{ms})$, with $\zeta$ being greater than unity.

Estimating $B_{H}$ is difficult. However, considering that the maximum rate at which the rotational energy from a black hole could be extracted does not exceed the Eddington luminosity (Znajek~\cite{znajek}), the magnetic field strength cannot be more than
\begin{equation}
B^{max}_{H}\sim 10^4 \left(\frac{M}{10^9M_{\odot}}\right) ^{-1/2}\: \textrm{G}\,. 
\end{equation}

Following Blandford~\cite{bland}, we assume that the poloidal magnetic field strength which threads the accretion disk varies as
\begin{equation}
B^{p}_{D}\propto r_*^{-n}\,,\; \textrm{where}\; 0<n<3\,, 
\end{equation} 
and the dimensionless radius is defined as $r_*=r/r_g$. 

Finally, the expression of the power of the jets driven by the accretion onto supermassive Kerr black hole, is the following:
\bea
P_{jets} & = & \dot{m}\dot{M}_{Edd}c^2\left( 1-q_m\right)\left[ E^{\dagger}\left( r_{sl*}\right) -E^{\dagger}\left( r_{ms*}\right)\right] + \frac{M^2G^2}{c^3}\cdot\frac{B_{H}^2\cos \theta}{\zeta}r_{ms*}^{n}(1+\sqrt{1-a_{*}^2}) \nonumber \\
&  & \int_{r_{ms*}}^{r_{sl*}}r_*^{3-n}\left[\frac{1+r_{*}^{-2}a_{*}^2+2r_{*}^{-3}a_{*}^2}{1-2r_{*}^{-1}+r_{*}^{-2}a_{*}^2}\right] ^{1/2} \left[\frac{a_{*}}{2(1+\sqrt{1-a_{*}^2})}-\frac{1}{r_{*}^{3/2}+a_{*}}\right]\frac{1}{r_{*}^{3/2}+a_{*}}dr_*.  
\label{Pjet} 
\eea
It shows that the jets power depends on the BH spin parameter, the accretion rate, the BH magnetic field strength, the ratio of the latter one to the poloidal magnetic field strength at the innermost stable orbit, the angle between the poloidal magnetic field lines and the accretion disk surface, and the scaling of the poloidal magnetic field with the accretion disk radius.

\section{Conclusion}
The amount of energy that the jets could receive from the accretion disk and black hole rotation depends on the details of the gas flow, the accretion rate, the black hole spin, and the magnetic field configuration that develop. Based on this fact, in the case of a thin accretion disk that accretes at the Eddington limit and is magnetically coupled to a highly spinning black hole ($10^{-6}\,\le \, 1 - a_* \, \le \; 0.05$) of $10^9\,M_{\odot}$, the jets power (Eq.~\ref{Pjet}) is roughly estimated to be $\sim 10^{45}\: \textrm{erg$\cdot$s}^{-1}$. Moreover, at low accretion rates, the jets can still be driven (see, e.g., Donea \& Biermann~\cite{db}, for a similar result). In this case, the efficiency of driving jets (defined as the ratio of the jets power to the rest energy of the accreting mass) can reach values larger than unity. In conclusion, this model provides an analytical approach to the jets driven by an accretion disk magnetically connected to a spinning black hole harboured in the center of an AGN.

\section*{Acknowledgments}
I.~D. thanks the organizers for partial financial support. Partial support for this work comes from the VIHKOS through the FZ Karlsruhe. I.~D. is now funded by the International Max Planck Research School (IMPRS).

\end{document}